\documentclass[conference]{IEEEtran}
\usepackage{subfigure}
\usepackage{moreverb}
\usepackage{epsfig}
\usepackage{amsmath,amssymb,amsthm,mathrsfs,amsfonts,dsfont}
\usepackage{adjustbox,lipsum}
\usepackage{algorithm,algorithmic}
\usepackage{amsfonts}
\usepackage{epsfig}
\usepackage{amssymb}
\usepackage{amsmath}
\usepackage{amsthm}
\usepackage{subfigure}
\usepackage{multirow}
\usepackage{rotating}
\usepackage{graphicx}
\usepackage{tabularx}
\usepackage{array}
\usepackage{anyfontsize}
\usepackage{color,soul}
\usepackage{graphicx,dblfloatfix}
\usepackage{epstopdf}
\usepackage{blindtext}
\usepackage{amsthm,amssymb,amsmath,bm}
\usepackage{subfigure}
\usepackage{amsfonts}
\usepackage{epsfig}
\usepackage{amssymb}
\usepackage{amsmath}
\usepackage{cite}
\usepackage{graphicx}
\usepackage{fancyhdr}
\usepackage{subfigure}
\usepackage[subfigure]{tocloft}
\usepackage[font={small}]{caption}
\usepackage{subfigure}
\usepackage{tabularx}
\usepackage{cite}
\usepackage{booktabs,multirow}
\usepackage{tabu}
\usepackage{longtable}
\usepackage[table,xcdraw]{xcolor}
\usepackage[english]{babel}
\usepackage[table]{xcolor}
\usepackage{listings}
\usepackage{comment}

\definecolor{tableHeader}{RGB}{211, 47, 47}
\definecolor{tableLineOne}{RGB}{245, 245, 245}
\definecolor{tableLineTwo}{RGB}{224, 224, 224}
\usepackage[utf8]{inputenc}

\definecolor{codegreen}{rgb}{0,0.6,0}
\definecolor{codegray}{rgb}{0.5,0.5,0.5}
\definecolor{codepurple}{rgb}{0.58,0,0.82}
\definecolor{backcolour}{rgb}{0.95,0.95,0.92}

\lstdefinestyle{mystyle}{
  backgroundcolor=\color{backcolour}, commentstyle=\color{codegreen},
  keywordstyle=\color{magenta},
  numberstyle=\tiny\color{codegray},
  stringstyle=\color{codepurple},
  basicstyle=\ttfamily\footnotesize,
  breakatwhitespace=false,         
  breaklines=true,                 
  captionpos=b,                    
  keepspaces=true,                 
  numbers=left,                    
  numbersep=5pt,                  
  showspaces=false,                
  showstringspaces=false,
  showtabs=false,                  
  tabsize=2
}

\begin{document}
\title{ Interaction and Conflict Management in AI-assisted Operational Control Loops in  6G }

\author{Saeedeh Parsaeefard, Pooyan Habibi, and Alberto Leon Garcia\\Electrical and  Computer Engineering Department, University of Toronto\\ {saeideh.fard, pooyan.habibi and alberto.leongarcia}@utoronto.ca}
\maketitle
\begin{abstract}
This paper studies autonomous and AI-assisted control loops (ACLs) in the next generation of wireless networks in the lens of multi-agent environments. We will study the diverse interactions and conflict management among these loops. We propose "interaction and conflict management" (ICM) modules to achieve coherent, consistent and interactions among these ACLs. We introduce three categories of ACLs based on their sizes, their cooperative and competitive behaviors, and their sharing of datasets and models. These categories help to introduce conflict resolution and interaction management mechanisms for ICM. Using  Kubernetes, we present an implementation of ICM to remove the conflicts in the scheduling and rescheduling of Pods for different ACLs in networks.  
\end{abstract}
\vspace{-.0cm}
\IEEEpeerreviewmaketitle
\section{Introduction}
The wireless community has been shifted its focus 5G to the next generation of wireless networks, e.g., 6G and 5G beyond \cite{9200631,8766143}, which will need to address AI and ML use cases. 6G is inherently a data-oriented and autonomous networking architecture where AI-assisted control loops (ACLs) reside inside the network to handle the operational aspects. Recently, standardization bodies have been active in defining how the networking should be modified and how the control loops can be sited inside of the networks. 

ETSI introduced the Experiential Networked Intelligence \cite{ENI} through the concept of closed control loops containing \textit{monitoring, analysis, policy, execution plus knowledge} steps, MAPE-K, with applications in infrastructure management, network operations, and service orchestration and management \cite{ENI2020}.  A service layer and network applications (Netapps) are introduce \cite{Viewon5GArchitecture} to provide open-access APIs to over the top (OTT) enterprises and facilitates instantiating of services through network slicing \cite{Viewon5GArchitecture}, and \cite{AIaaSpaper} introduces the notion of AI-as-a-Service being offered by network providers. In this work, ACLs are modeled as a chain of computing, storage, and communication service functions for OTT and networking applications. Like slices, ACLs can be represented as a graph of network functions (NF) and ML-oriented service functions for AI-based use cases (or slices) inside the network. However, there has been no discussion the autonomous and automatic behaviors needed in ACLs, and there have been no systematic discussions about how ACLs will interact with each other. In this paper, we address these two aspects of ACLs in networks. 

\begin{figure*}
	\begin{center}
		\includegraphics[width=6.7 in]{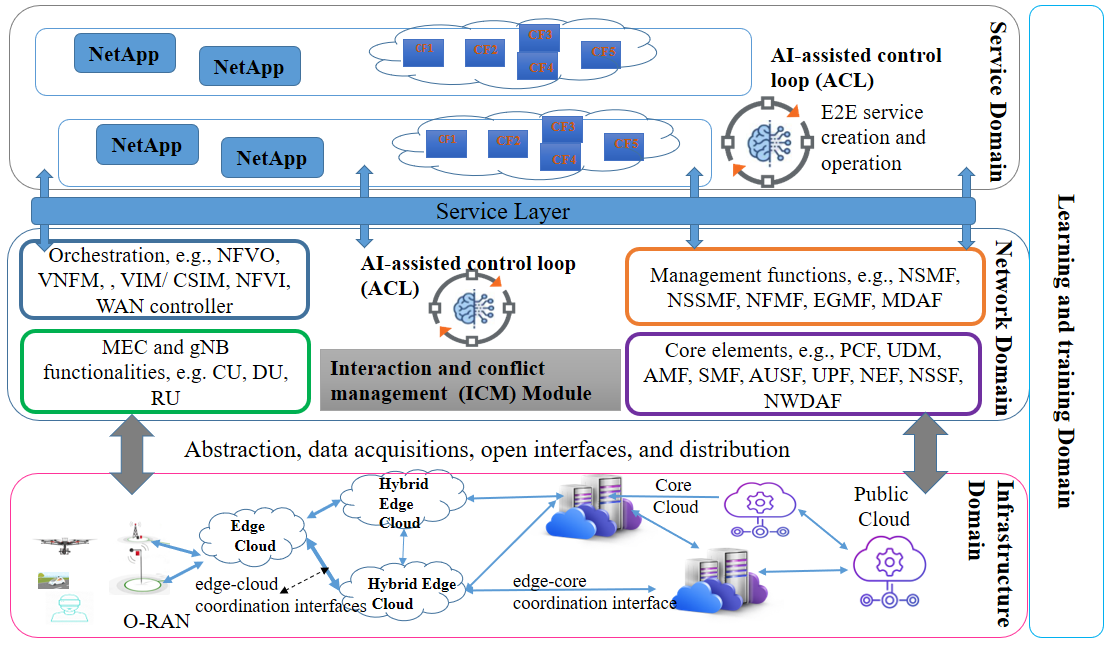}
		\caption{6G Domains for Future Wireless Networks}
		\label{fig1}
	\end{center}
\end{figure*} 
The autonomous behaviors of ACLs will change the nature of 6G and transform it into a multi-agent environment. Here, the control and management of tasks of the networks will be handed over to ACLs which will need to deal with diverse and conflicting objectives, e.g., energy efficiency versus resource utilization. ACLs can either cooperate to schedule networking procedures and to share resources or they can compete for limited physical and virtual 6G resources. They can capture their lessons learned from training or testing phases as knowledge they that they share with each other, their datasets, and models \cite{9388790,9328851,DBLP:journals/corr/abs-2102-07572,AIaaTL}. When the network management is unlocked from rigid, tight procedures to a more multi-agent dynamic environment, it becomes necessary that the interaction and conflict control be consistent and that network performance be coherent. In this paper, our goal is to address this challenge. 

To realize interaction and conflict, we resort MAPE-K loops to realize ACLs in a more systematic way. We view ACLs as autonomous agents in the networks. We identify different categories of interactions among ACLs based on their cooperative or competitive behavior, transferring knowledge among the first two steps of MAPE-K loops. We also use the special-temporal structure of the wireless networks and the size of the control parameters of ACLs, and propose to have Femto, Micro, Macro, and Mega ACLs definitions. These categories lead to characterize functionalities of the \textit{interaction and conflict management} (ICM) unit inside 6G. We also introduce the new network procedures for handling these interactions. Finally, we provide a realization of ICM using existing platforms as follows. We consider scheduling among parallel and competing ACLs for placing Pods on computing resources and we use Kubernetes \cite{dobies2020kubernetes} to resolve conflicts of ACLs. The organization of this paper is as follows. In Section II, we study domains of 6G. In Section III, interactions and conflicts among ACLs are investigated, followed by discussions about the ICM module in Section IV. Section V involves implementation use cases over Kubernetes for ICM module; followed by conclusions and future research in Section VII.
\section{ 6G Architecture Domains}

Fig. \ref{fig1} presents the 6G domains and its new features. The 6G architecture will extend 5G in many aspects such as  the infrastructure and abstract layers. In the bottom, we have a multi-tier cloud-native network extending from the edge to the core and comprising both computing, processing, storage, networking, and service functions in virtualized and containerized form. The cloud structure has hybrid (private and public) and heterogeneous structures from the edge to the core. Edge clouds may be connected to front edge transceivers in different frequency bands with software-defined radio structures. In the core of 6G, computing can be provided by different clouds, including public clouds (e.g., Azur and GCP), infrastructure clouds (e.g., edge and core of a service provider), and mixture of these two groups \cite{HexaX}. 
These heterogeneous and hybrid cloud features required that we have a more unified view of transport layers among different parts of the network where their different transmission links need to be coordinated. This hybrid and heterogeneous end-to-end virtualized cloud environment provides a pool of storage, computing, and communication resources, that can be viewed as an abstracted graph of resources on top of the infrastructure layer. The integration of virtualized and software-defined structures can also unlock access to any data pass through interfaces, physical and virtualized entities inside the networks. More importantly, any NF and service functions can have similar features with the help of abstraction through the software-defined and virtualized structure. Therefore network domain and service domain functions can be seen as a graph of cloud-native functions embedding in the networks. In the network domain, we provide functions related to the following parts of the network 
\begin{itemize}
    \item Management entities 
    \item Orchestration entities
    \item 5G access and mobile edge computing (MEC) functions 
    \item Core functions.
\end{itemize}
Above the network domain, there is a service layer which is a common interface that enables the interaction between the service intelligence and the underlying network. This layer provides north-bound interfaces for over-the-top applications for exposure of the network management, core and access elements. The service layer is responsible for providing "instantiation of slices for each enterprise," "Orchestration of application-layer virtualized functions," "Monitoring and runtime management," and "operating the slice" for each OTT application. The service layer also takes care of all low-level functionalities for the slice owner and provides a trustful environment for their network usage by a specific slice \cite{Viewon5GArchitecture}. 

At the top of the network, there is a service domain containing two main elements:
\begin{itemize}
    \item Slice  function chain: a graph of network functions to serve the slice based on the requested QoS and the SLAs
    \item Network Application (NetApp): a main customer North-bound APIs  which is a software responsible for interacting with the control plane of a mobile network by exposed APIs and for composing services for vertical industries
\end{itemize}
The slice chains will be embedded inside the networks depending on their quality of service (QoS) requests. The edge clouds can provide very low latency computation and storage facilities for the slices. However, they suffer from limited computation capacity and higher cost. Core clouds provide a higher and more cost-effective capacity for both computation and storage facilities. However, they induce more E2E delay per slice. 

There are two main ACLs inside of this architecture. The first one belongs to the service layer for E2E service creation and operation, and the second one provides the operator control loop for the network management domain's functionalities. While there are two different domains, they are inherently interrelated in providing the services and slices, and therefore, they are not standalone. 

In addition to fully automated network management, the network domain should determine and resolve all different interactions and conflicts among the ACLs to have consistent and coherent network automation. We underscore the importance of this point by introducing the box behind the ACL of the network domain. This box should include all functions related to verifying the accuracy, coherency, and consistency of the output of ACLs, their conflicts, severe damage, and required interactions. We call this box interaction and conflict management (ICM), and we will discuss the nature of the functions of this box later. ICM has new procedures to handle the priorities of ACLs, stability analysis of network and other systems when more than one parallel ACLs are run simultaneously. It also initiates the request for transfer of knowledge among parallel ACLs with the help of the learning and training domain in 6G.

The learning and training domain is a new domain in the networking architecture to integrate AI and ML Training phases into the networking environment. This domain has different sandboxes of training of ACLs, their retraining, and examining any feature in offline, near online, and online manners. Furthermore, this plane should model the interactions between ACLs to achieve ICM. Also, there are sandboxes to emulate the interaction and conflict situations among ACLs to anticipate and prevent undesirable scenarios. Also, this plane provides the repository of models and datasets for ACLs to transfer their knowledge and have more interactions during the training phases \cite{AIaaSpaper,AIaaTL}.

\section{Interactions and Conflicts among ACLs} 
\begin{figure}[ht]
\begin{center}
\includegraphics[width=0.34\textwidth]{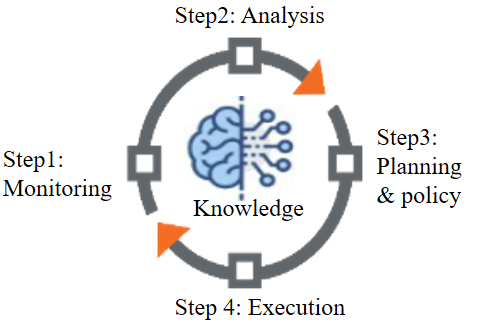}
\end{center}
\caption{ ACL in 6G from the concept of the MAPE-K loop}\label{fig2}
\end{figure}

As the first step to realize ACLs in 6G, we resort to the concept of MAPE-K loops \cite{ENI, 120891009, AIaaSpaper} as depicted in Fig. \ref{fig2}. There are four major steps in MAPE-K loops as follows:
\begin{itemize}
    \item Step 1 is a \textit{m}onitoring phase, including the data gathering, which consists of monitoring an environment via sensor devices, measurement tools and collecting data within a suitable time window. For this paper, the environment is the physical and virtual entities inside of 6G, and the measurements may include the traffic of users, QoS, and network resource and entities states;
\item Step 2 is a data \textit{a}nalyzing step involving diverse ML and
AI-based algorithms to train a specific model based on the collected data of step 1 to realize the pattern in datasets;
\item  Step 3 is about policy making of step 2, including
transformation of the results of step 2 to parameters understandable by the network and the time for an update or taking action according to the results of step 2
\item Step 4 is about implementing actions inside of the networks autonomously or by human intervention. The action will be implemented in the related domains, e.g., network or service, or infrastructure domain.
\end{itemize}
Knowledge is the essence of this iterative learning loop which can be extracted from data-oriented algorithms.
\begin{figure*}[ht]
\begin{center}
\includegraphics[width=.70\textwidth]{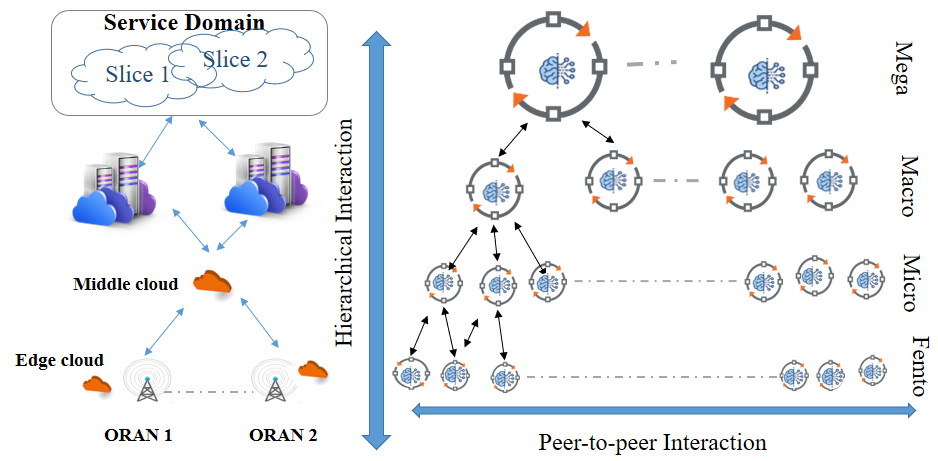}
\end{center}
\caption{Categories of ACLs based on their sizes and spatio-temporal features  }\label{fig3}
\end{figure*}

To address the conflict and interaction among ACLs, consider three following examples for service and network domains:

\begin{itemize}
    \item Example 1: Resource management ACL for RAN in network domain to control the size and the placement of NFs inside of the clouds of RAN;
    \item Example 2: Resource management ACL for Core in network domain to control the size and the placement of NFs inside of the clouds of core;
    \item Example 3: ACLs for E2E slice instantiating.
\end{itemize}

The ACL of example 1 needs to monitor the traffic of users and the state of the available resources per each cloud of RAN. For traffic monitoring, the historical data and the measurements of all ingress nodes of the access and requests of the end-users can be collected in Step 1 and then analyzed in Step 2. In step 2, a trained model based on a historical dataset predicts the required resources and places the containers related to required NFs based on the traffic variations. Step 3 makes policies and commands for the edge clouds to run, scale up or down, or terminate NFs in the network. After taking actions by edge-cloud, the resource states of the edge will be updated. The ACL of example 2 also requires traffic prediction and network state collection. The traffic prediction of ACL of example 1 can help ACL of example 2 predict part of the network's traffic, or they can share their data sets. In step 2, the trained model of ACL predicts the resources. The model of this example has a different spatio-temporal traffic granularity compared to the model of example 1. However, both models have some hierarchical similarities due to the hierarchical and spatio-temporal nature of wireless networks \cite{AIaaTL}. In steps 3 and 4, the action to scale up or down, instantiation, and terminations of containers of core cloud related to NFs will be handled. The state of the network after that should be updated. In example 3, based on the request of OTT providers, one slice should be initiated inside of the network. For this example, the ACL needs all the states of network entities, available resources, the traffic estimation of other slices, and the SLAs and QoS of existing and new slices to decide the initiating of E2E slice instantiating. The data collected from examples 1 and 2 can be used here. In Step 2, more data should be ingested at the training phase to train a more complex model than models of examples  1 and 2. 

\begin{figure*}
	\begin{center}
		\includegraphics[width=5.7 in]{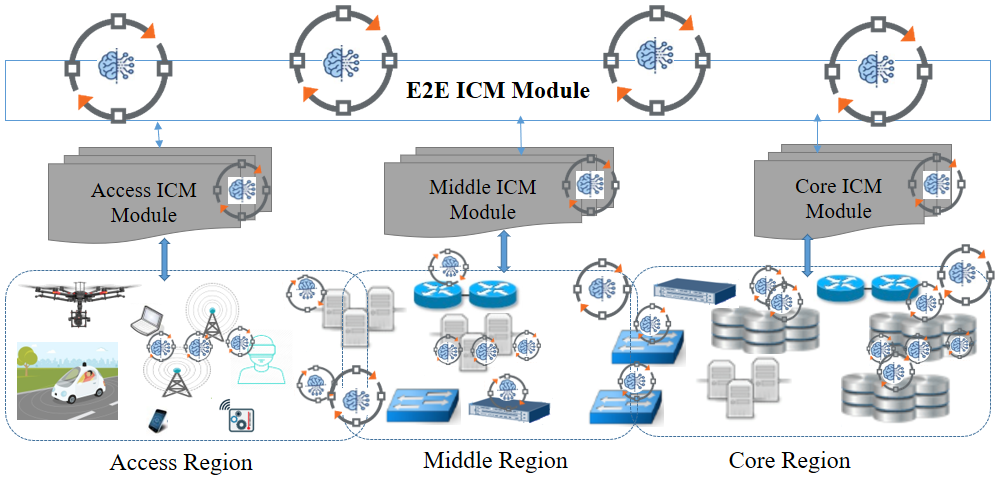}
		\caption{ Multi-tier and recursive structure of ICM in 6G }
		\label{fig5}
	\end{center}
\end{figure*}

These three examples clearly show some overlaps in Step 1 and Step 2, e.g., the dataset of example 1 from the Step 1 can be shared among the other two examples. The model of example 1 can also help model 2 of example 2. However, the actions and the policy of these three examples can conflict. For example, Examples 1 and 2 aim to scale down their container size due to the slow decrements of existing traffic; but the ACL of Example 3 increases the size of the container due to the new instantiating of one slice and extra traffic prediction. 
The above observations from these three examples show that the ACLs are not standalone in different domains of 6G, and inherently they have a lot of interactions, communications, and conflicts. Therefore, we consider each ACL as one autonomous agent that resides in the network, and ACLs' interactions and overlaps with others can be seen through the lens of a multi-agent environment. Therefore, 5G B and 6G can be considered an environment in the presence of intelligent agents aiming to control the resources based on their objectives. Similar to any multi-agent environment, managing this autonomous interaction will be one of the main challenges. Also, they can be in diverse modes of interactions in this environment: e.g., cooperative versus competitive, isolated versus overlapped agents. Therefore, we aim to provide comprehensive classifications that can help to realize these interactions and conflicts. We use two approaches for this goal. We first use the steps of MAPE-K loop to make two categories of ACLs. Then, we use the spatio-temporal features of wireless networks to reach the third category among the ACLs. 
\begin{itemize}
  \item  \textbf{Category 1}: In each ACL, steps 1 and 2 and knowledge are related to the data science and machine learning algorithms. Therefore, interactions in Step 1 to share the data, in step 2 to share and reuse the models, and transferring the knowledge among ACLs can be handled by the concepts of multi-task learning, transfer learning, and meta-learning \cite{AIaaTL}. In this context, usually, there is a source ACL that shares its dataset or its model parameters with another or a set of ACLs, called targets. Category 1 shows if the ACLs share their datasets and models or not, called the interaction in the training phase among ACLs.  
        \item \textbf{Category 2}: Step 3 and Step 4 are related to applying a policy, the time for taking actions, and the results of actions in this environment. In this case, some actions can conflict, as we show for example 3 and the outputs of examples 1 and 2. In this sense, ACLs can be competitive or cooperative. For example, they can compete to utilize one resource or cooperate in running one similar action for their own tasks. 
       \item \textbf{Category 3}: As depicted in Fig. \ref{fig5}, we use the spatio-temporal features of the wireless networks of classifications among ACLs as follows: 
    \begin{itemize}
    \item Femto ACL: Tiny ACL for a specific container related to one specific network function (NF), e.g., scale up or down of one container inside of one cloud
    \item Micro ACL: Small but isolated ACL in one virtual or physical entity inside of the network, e.g., scheduling of resources for one cloud inside of the network
    \item Macro ACL: A medium-size ACL which responsible for the functionalities among a set of entities belonging to one region, e.g., ACL for resource allocation among the hybrid edge clouds in one region 
    \item Mega ACL: An E2E ACL over the service layer or related the slice's functionalities, e.g., E2E service creation and operation, slice isolation control and resource allocation, and slice scheduling
\end{itemize}
\end{itemize}

From Femto to Mega ACLs, we can see that ACL output can impact more entities and NFs inside of the networks. Therefore, the output of one Mega ACL can impact the large numbers of Micro, Micro, and Femto ACLs. However, the output of one Femto ACL can just change the states and the environment for some other Femto ACL and its related Macro and Micro ACLs. This explanation can show the inherent feedback loops among all ACLs inside of the networks where the stability of the network is related to how we control the feedbacks among the ACLs and how we check the coherency of the outputs of ACLs to be sure that there are instabilities among these feedback loops. 

For example, implementing the output action of one Mega ACL can change the states of many other ACLs from Mega to Femto ACLs. All of these ACLs need to change their outputs according to the new network states. However, on the other side, even the output of one Femto ACL can change the environment of other ACLs, but the impact of Femto ACL over the network states is more localized and isolated than the Mega ACL which has more global impact overall network states. 

Indeed, Category 3, along with Categories 1 and 2, can help us better realize interactions among ACLs. However, the network management should be equipped with the mechanism to control interactions and to prevent the instability of the network. In Fig. \ref{fig3}, in addition to hierarchical interactions, we also highlight the peer-to-peer interactions among the same groups of ACLs in Category 3. These peer-to-peer interactions contain cooperation and competitive models, and knowledge sharing among the same peers in the network. Furthermore, spatio-temporal features of wireless networks determine the interactions among ACLs and can show the historical behavior of the network per each level and region. In the following sections, we will discuss how these features, along with other approaches, can help prevent instability of the network and conflict among ACLs and manage their interactions. 
 
\section{Interaction and Conflict Management (ICM) Module in 6G}
The above categories and examples highlight that the management in 6G deals with many new aspects of managing the dynamic environment among the ACLs, which can have different types of interactions, competitions, or cooperation, and finally, they can cause instabilities. Therefore, we introduce the ICM module for 6G management functionalities. ICM is responsible for providing appropriate procedures to handle all these conflicts and interactions to guarantee the stability of the network under the autonomous behaviors of ACLs. Based on the categories of ACLs, ICM should provide the following mechanisms: 
\begin{itemize}
    \item A mechanism for exchanging the datasets and models for Category 1: This can include providing a trustworthy list per each ACL, containing the information about other ACLs allowed to get the datasets or models. This list should contain the required security approaches per each exchange of information.
    \item For Category 2, the conflict control procedure and conflict resolve algorithm to access the resource should be provided among ACLs. One approach to provide them is to make priorities among the ACLs and provide appropriate scheduling among the ACLs to utilize the resources. Also, any back and forth actions causing interference among the actions of ACLs should be prevented here, e.g., when the energy efficiency module turn-off one cloud while the ACL for load balancing module turns it on. 
\item For Category 3, a coherency check of the outcome of ACLs based on the available historical and spatio-temporal features should be provided. In addition, any anomaly outcome of one ACL compared to the historical view of the network should be prevented and isolated inside of the network to prevent unpredictable situations inside of the network. 
In case that some ACLs have very abnormal behaviors, e.g., their output actions are far from expected actions of the history of the networks, those ACLs can be suspended or put under observations, and third parties or operators can check their outputs. Consequently, the stability of the network can be guaranteed. 
\end{itemize}

The final question is if the ICM module should be implemented in centralized or decentralized scenarios. While there is always a trade-off between centralized and distributed approaches, scalability issues, hierarchical features, and spatio-temporal aspects of wireless networks lead to distributed but hierarchical implementation. One presentation of this approach is shown in Fig. \ref{fig5} where per each region pf 6G, there exists one ICM module and on the top one E2E module provides integration among the decision making procedure. The E2E ICM module can also handle the time granularity and the interaction among the ICM modules of regions. This implementation is close to the recursive-hierarchical structure of ETSI MANO \cite{5Gvision} and can integrate inside 6G and 5G beyond straightforwardly.


\section{Implementation of ICM by Kubernetes}
\begin{figure*}
	\begin{center}
		\includegraphics[width=5.9 in]{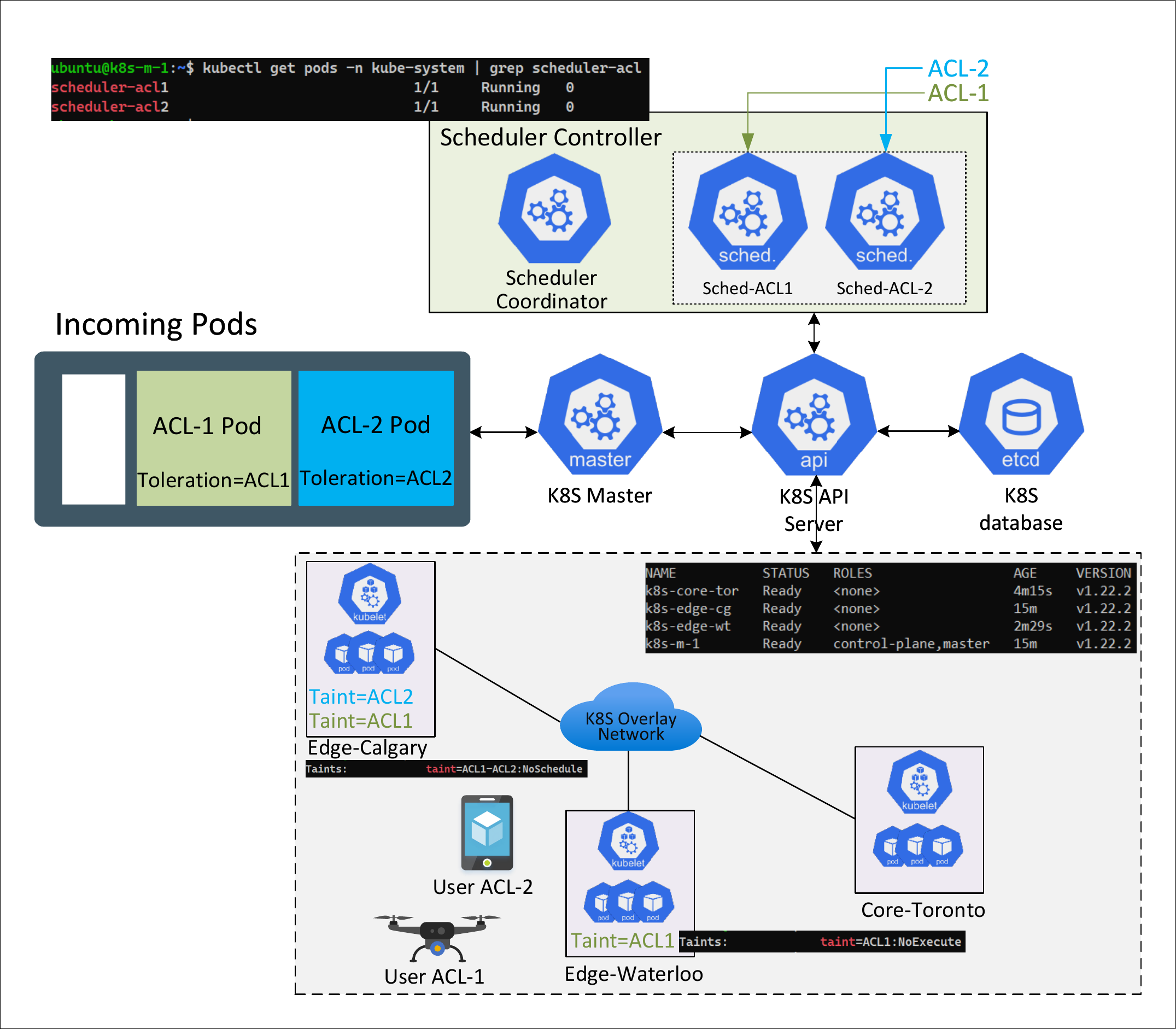}
		\caption{ICM prototype deployment using Kubernetes on SAVI}
		\label{fig:Sched-multi}
	\end{center}
\end{figure*}

In this section, we prototype three cases for ICM using the Kubernetes scheduling features. For all cases, we use Kubernetes clusters and implement ACLs on the SAVI testbed \cite{Kang2013May}. Consider two parallel ACLs responsible for running NFs of two slices, where ACL1 has a higher priority than ACL2. There are new users for these two ACLs, and ACLs need to run new NFs in the proximity of the users over the cloud edges, as depicted in Fig. \ref{fig:Sched-multi}. Cloud edges are resource-limited, and there is a chance there may not be enough available resources for both ACLs. The output of each ACL is run by one scheduler in the network. Now our goal is to show how Kubernetes can help to provide a new method to have the conflict management by ICM. In Kubernetes, we have three major elements \cite{dobies2020kubernetes}: 
\begin{itemize}
    \item Pods are units of resources to run a workload by placing containers on Nodes. 
    \item Nodes may be virtual or physical machines, depending on the cluster. Each node is managed by the control plane and contains the resources necessary to run Pods. We have deployed a cluster including three connected nodes (datacenter) in Fig. \ref{fig:Sched-multi}, one node as a core located in Toronto, and two edge nodes in Calgary and Waterloo. Our edge nodes have fewer resources compared to the core node. The components on a node include the Kubelet, a container runtime, and the Kube-proxy. Pods in the cluster are connected through the main virtual overlay network by the Calico network plugin.  
    \item The controller scheduler contains kube-schedules and scheduler coordinator in the control plane and assigns Pods to nodes. The scheduler determines which nodes are valid for the placement of each Pod in the scheduling queue according to constraints and available resources.  By default, Kubernetes has one scheduler which ranks each valid node and binds the Pod to a suitable Node. 
\end{itemize}


\begin{lstlisting}[language=Python, caption=Priority class for ACL1, label={lst:priorityandpreemting}]
apiVersion: scheduling.k8s.io/v1
kind: PriorityClass
metadata:
  name: ACL1
  label: sched-ACL1
value: 10
preemptionPolicy: Yes
globalDefault: false

"This priority class is defined for ACL1 Pods and can preempt any lower priority classes."
\end{lstlisting}


\begin{lstlisting}[language=Python, caption=Priority class ACL2, label={lst:priorityandpreemting2}]
apiVersion: scheduling.k8s.io/v1
kind: PriorityClass
metadata:
    name: ACL2
    label: sched-ACL2
value: 5
preemptionPolicy: Yes
globalDefault: false

"This priority class is defined for ACL1 Pods and can preempt lower priority classes."
\end{lstlisting}


Here, we modify these elements based on Kubernetes features to introduce new attributes to control the conflicts as follows:
\begin{itemize}
    \item  As shown in \ref{fig:Sched-multi}, per each node, we add "taint" as a new attribute. If the taint value of one node is equal to ACL1, e.g., edge-waterloo, it means ACL1 has a high priority to run its own Pod at this node. Each node can have multiple taints, e.g., edge-Calgary. We can run each node in three taint modes as follows:
        \begin{itemize}
            \item \textit{NoSchedule}: NoSchedule: if this taint is applied to a node that contains some Pods that do not tolerate this taint, the node does not exclude these Pods. But no more Pods are scheduled on this node if they do not match all the taints of this node.
            \item \textit{PreferNoSchedule}: Similar to \textit{NoSchedule}, this taint may not allow Pods to be scheduled on the node. But this time, if the Pod tolerates one taint, it can be scheduled. As shown in \ref{fig:Sched-multi}, the node in edge Calgary has a "taint" as this mode and will be a suitable host for Pods from ACL1 and ACL2.
            \item \textit{NoExecute} applies to a node excluding all actual running Pods which are not able to be hosted by the node. These Pods will be evicted and should be rescheduled on another node. As an example, in Fig. \ref{fig:Sched-multi} we taint the edge in the Waterloo for ACL1.
        \end{itemize}
    \item Per each Pod, we define a toleration and the priority class to which they belong.
    \item Per each ACL, we define one schedule unit in the controller scheduler, responsible for running the Pod of ACL on the node. ACLs are implemented as separated Pods in "kube-system namespace" as depicted in Fig. \ref{fig:Sched-multi}. Each scheduler is assigned with one priority level value, and a larger value means a  higher priority. We present the priorities of ACL1 and ACL2 as 10 and 5, respectively, and we show them in Listing \ref{lst:priorityandpreemting} and Listing \ref{lst:priorityandpreemting2}.
\end{itemize}

\subsection{Case 1}
\begin{figure}[ht]
\begin{center}
\includegraphics[width=0.49\textwidth]{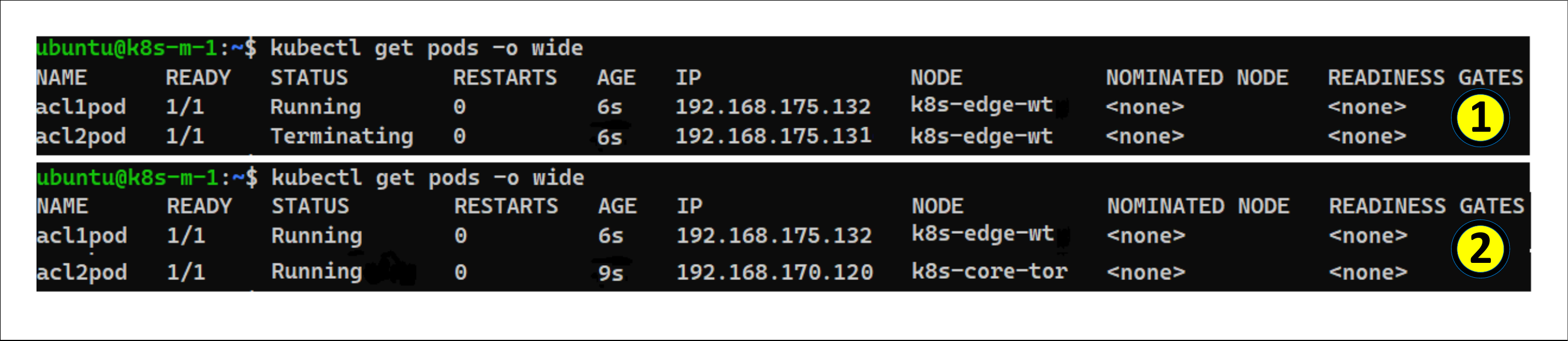}
\end{center}
\caption{Output of placement of ACl 1 and ACL 2 in edge Waterloo for the Case 1}
\label{fig:Case1-Result}
\end{figure}

We run two parallel ACLs, which are responsible for running two Pods of two different slices simultaneously. Assume the outputs of both ACLs are to place their Pods in edge cloud Waterloo, which has the capacity to run one of these Pods. To resolve the conflict of using the resource of this edge server,  the value of $ACL$ priority ($ACL_P1$) to have priority mechanism among the ACLs. For example, in this case, $ACL1_P 1=10$ and $ACL_P 2=5$ and since $ACL_P1 > ACL_P2$, ACL1 has a priority to use the resources of the edge cloud Waterloo.  Therefore,  the coordinator in Kubernetes selects ACL1 scheduler for placing ACL-1 Pod on the edge-Waterloo, and ACL2 should be rescheduled to another cloud either on the edge-Calgary or core-Toronto. Listing \ref{lst:priorityandpreemting2} shows the priority classes for ACL 1 and ACL 2. Fig. \ref{fig:Case1-Result} shows the results of applying two Pods from ACL1 and ACL2 on edge Waterloo, which the Pod from ACL1 goes in the running state and the Pod from ACL2 does not allow to run on this node and it is rescheduled on the other node which in this case it is core Toronto. 

\begin{figure*}
\begin{center}
\includegraphics[width=0.76\textwidth]{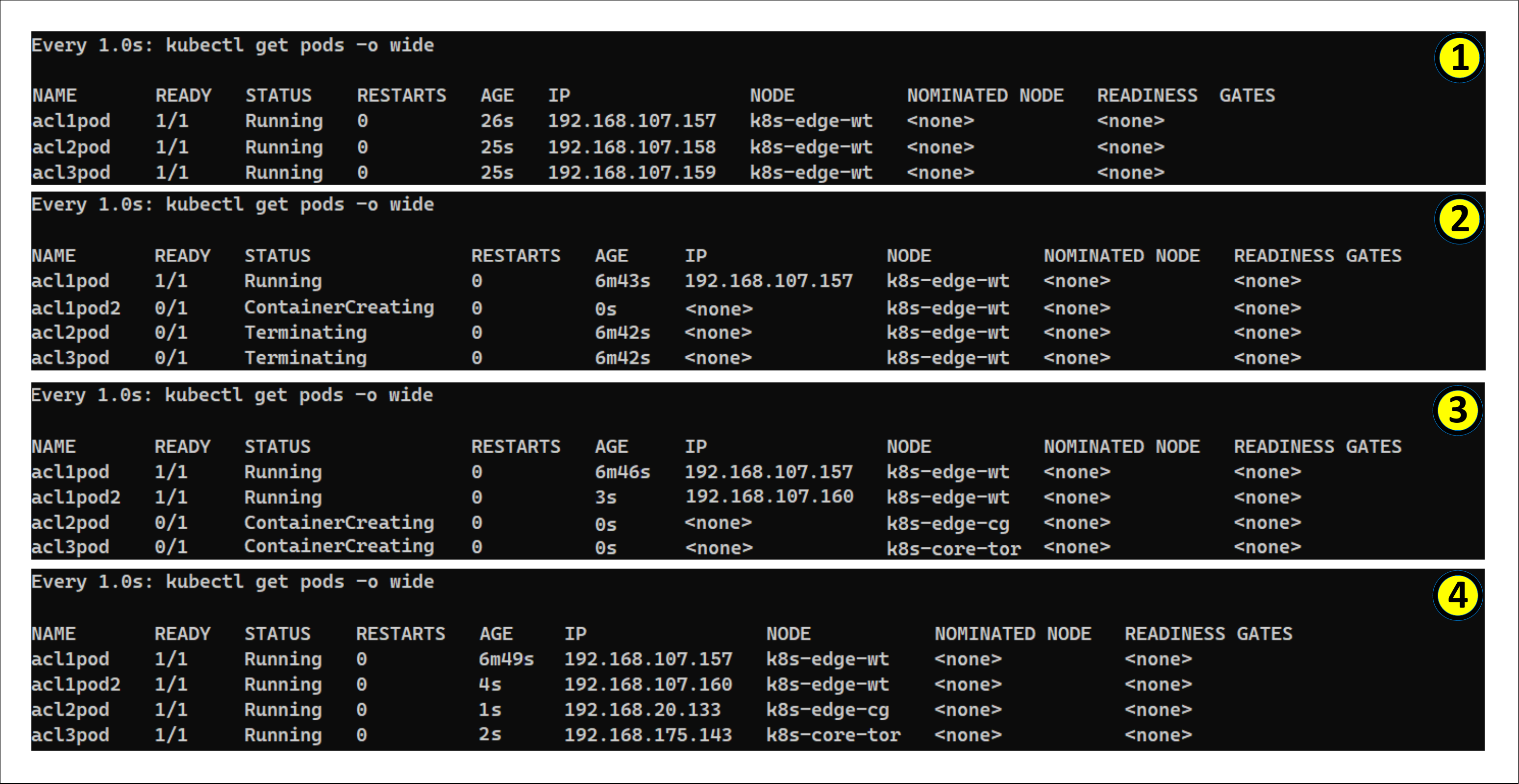}
\end{center}
\caption{Output of placement of ACl 1 and ACL 2 and ACL3 on SAVI nodes for the Case 2}
\label{fig:Case2-Result}
\end{figure*}

\subsection{Case 2}
Here, there are running Pods from low-priority ACLs on edge Waterloo, e.g., ACL2 Pods and ACL3 Pods, which are shown in Fig. \ref{fig:Case2-Result}. Now, ACL 1  aims to run its Pod, which should be placed on edge Waterloo. Since the schedule of this new ACL has a high priority compared to the other existing ACLs of Pods, the coordinator chooses ACL1 to run its Pod on edge Waterloo. However, edge Waterloo does not have enough resources. So, a conflict among ACLs' Pods will occur that at the end, the ACL1 should be able to run its Pods on edge Waterloo, and the low priority pods should be terminated and rescheduled on the other nodes available. To do so, we use taint and toleration to implement this use case. As it is shown in the \ref{fig:Case2-Result} the edge Waterloo has been tainted as configured as a "NoExecuted" to be a priority host for ACL1 pods, other ACLs can place their pods on edge Waterloo as long as it has enough resources for ACL1 pods. Otherwise, the pods from other ACLs will be evicted to free up space for ACL1 pods. Edge Calgary has been tainted as a priority host for both ACL1 and ACL2 pods; however, it performs the priority for new ACL1 and ACL2 pods and does not terminate the current running pods from other ACLs to free up space for ACL1 and ACL2. The core Toronto does not taint, and it hosts all ACLs pods. In Fig. \ref{fig:Case2-Result} (1), we show the current status of running Pods on edge Waterloo. Then, in Fig. \ref{fig:Case2-Result} (2), ACL 1 asks for a new Pod, e.g., ACLPod1, that should be placed on edge Waterloo. Since this edge does not have enough memory space to run this Pod, in Fig. \ref{fig:Case2-Result} (3), we show how by our setting, Pods from ACL2 and ACL3 are evicted from the edge Waterloo to provide space for running ACLPod1. Then these Pods are rescheduled by ACL2 and ACL3 as shown in Fig. \ref{fig:Case2-Result} (4) and replaced in edge Calgary and core Toronto nodes, respectively.

\section{Conclusion }
In this paper, we address the conflict and interaction management among AI-assisted operational control loops (ACLs) in future wireless networks (6G). We discuss how the autonomous behavior of ACLs changes the nature of wireless networks into multi-agent systems. Therefore, controlling and managing their interactions are essential parts of 6G. We discuss several categories of ACLs to fully understand the impact of ACLs' interactions. Then, we introduce mechanisms in the new module ICM in 6G to handle the conflicts and interactions. We demonstrate how attributes in Kubernetes can be changed to reach these goals. We believe this paperis a first step to realize the diverse aspects of these new requirements of 6G, and there remains a lot of room for research and proposing mechanisms to handle the interactions of ACLs. 
\bibliographystyle{IEEEtran}

\bibliography{IEEEabrv,myref}

\end{document}